# One-way Valley-locked waveguide with large channel achieved by all-dielectric Photonic Crystals

Li Liang[1], Xiao Zhang[1], Chuan Wang[2], Jie Liu[1], Longzhen Fan[1], Chengpeng Liang[1], Liang Liang[1], Feifei Li [3]*, Qi Wu[4]*, Yin Poo[1]*

*1 School of Electronic Science and Engineering, Nanjing University, Nanjing 210093, China*

*2 The General Design Institute of Hubei Aerospace Technology Research Institute, Wuhan, China*

*3 School of Information and Control Engineering, China University of Mining and Technology, Xuzhou, China*

*4 School of Electronics and Information Engineering, Beihang University, Beijing, China*

[*]Correspondence and requests for materials should be addressed to feifeili@cumt.edu.cn (Feifei Li), qwu@buaa.edu.cn (Qi Wu), ypoo@nju.edu.cn (Yin Poo)

**Nonreciprocity, which denotes the asymmetric or even unidirectional transmission of light, constitutes the cornerstone of modern photonic circuits. In the realm of photonic devices, it has been widely utilized in isolators, circulators and so on. Recent topology in artificial materials, an unprecedented degree of freedom, has been proposed to solve the effect of impurities on nonreciprocal transmission. However, in view of the bulk-edge correspondence, the spatial width of the transmission channel with uniform field distribution is quite narrow and needs further exploration. In this paper, we proposed a one-way valley-locked waveguide with a large channel in an all-dielectric photonic crystal. Quite different from the topological edge modes, the unidirectional property of our waveguide comes from the bulk modes with valley-lock, which can fully utilize the whole dimension of the structure with an efficiency of 100%. Additionally, the electrical field is uniformly distributed across the entire channel, which opens a**

**new avenue for low-loss nonreciprocity devices.**

# I. INTRODUCTION

As early as 1530 A. D., a marvelous piece of architecture named the Echo Wall was completed. The architecture facilitates clear auditory communication between individuals positioned on either side, once no obstacles surrounded the wall and people facing in particular directions [1]. This astounding phenomenon is known as reciprocity, a ubiquitous property in wave systems adhering to Lorentz reciprocity principles, and has been widely used in modern information fields such as antenna systems [2]. Reciprocity exists under strict conditions, akin to the case in the echo wall. Any alteration in the conditions leads to non-reciprocity. In the realm of communication, non-reciprocity, especially unidirectional transmission, plays a pivotal role in enhancing the noise immunity of signal processing systems, which can effectively isolate noise and echo from transmitted signals [3–8]. Traditional methods for non-reciprocal light transmission, like ferrite isolators and circulators [8–15], often rely on magneto-optical materials whose magneto-optical responses exist only at microwave frequencies and vanishes at higher frequencies, limiting further application at visible wavelength range. Recent advancements in photonic topological states have emerged as promising avenues for one-way propagation of electromagnetic (EM) waves with strong robustness to imperfections [9–11,16–21]. Nonetheless, as depicted in Fig. 1(a), the electric field of topological edge states is primarily distributed around the boundary of the topological insulator (TI). The one-way transmission channel is significantly narrower transmission channel compared to the entire volume of the TI, impeding further practical applications. Such low spatial utilization stems from the bulk-edge correspondence [22], which requires a sufficiently large volume for the nontrivial bulk phase and perfect topological bandgap. Therefore, the realization of a unidirectional waveguide capable of wide transmission channels and portable to arbitrary frequency bands becomes paramount for various optical applications.

Recently, an acoustic nonreciprocal transmission, featuring a broad channel induced by rigid acoustic boundary conditions [23], has been proposed. Compared to the topological edge states, trivial valley-locked bulk modes have been adopted to expand the transmission channel, providing a promising platform for one-way waveguides. However, to date, analogous outcomes in photonic systems remain unexplored.

In this work, we proposed a novel one-way waveguide with a wide channel by all low-loss dielectric photonic crystal (PhC). Valley-locked bulk states have been built to make the full utilization of the structure volume as illustrated in Fig. 1(b). Our approach is to eliminate one pair of bulk modes (blue line in Fig. 1(c)) and retain the other pair with opposite parities depicted as the orange line in 1(d). EM waves with different circular polarizations at orange modes in two different directions propagate along opposite directions, forming the one-way property. This one-way waveguide exhibits a wide transmission channel with a remarkably uniform of field distribution.

## II. MODELS AND THEORY

We first established two pairs of bulk modes with opposite parity along opposite propagation directions. As depicted in Fig. 2(a), the designed two-dimensional (2D) PhC with hexagonal lattice ($a$ = 16mm) is composed of dielectric rods ($\varepsilon$ = 12.4 and r = 2 mm) placed in the air. Under the $C_6$ inversion symmetry and time-reversal symmetry, two bulk bands (orange and blue) intersect at K and K' points in Fig. 2(b), known as the Dirac points. From the Poynting vectors and eigen electric field distributions in right panel of Fig. 2(c), we can infer two important properties. One is the orange and blue bands belong to the odd and even parity bands, respectively, with antisymmetric and symmetric electric field distributions along the mirror axis (gray line in Fig. 2(c)), which indicates one of the two curves can be selectively eliminated by modifying the boundary condition. The other is both blue and orange bands have opposite parities around K and K' points, i.e. valley-locked [19]. Focused on the lower half of the unit cell below the mirror axis, the vortex of energy flow is anticlockwise at K' point and clockwise at K point for the orange band, compared to clockwise direction at K point and anticlockwise direction at K' point for the blue one. Given that Γ - K and Γ- K' correspond to two opposite directions, the one-way bulk modes with valley-lock are naturally realized once either the orange or blue band is eliminated, corresponding to EM waves with opposite circular polarization will propagate in opposite directions.

Then, we attempted to eliminate one band to achieve the one-way bulk modes. Following the principle of mirror symmetry, the mirror axis can emulate a perfect electric (magnetic) conductor for the orange (blue) band. Setting the perfect electric conductor (PEC) boundary condition along the mirror axis to break modes' symmetry, the even modes (blue band) will disappear while the odd modes (orange band) remain. Likewise, the perfect magnetic conductor (PMC) boundary condition leads to the vanishing of the odd modes. In this work, we adopted the PEC boundary to achieve mode selection for its easy implementation. As depicted in Fig. 2(d), a pair of parallel metallic slabs (marked by red lines) served to truncate the infinite PhC into a semi-infinite ribbon-like structure with semi-rods along both the upper and lower boundaries. The projected band structure (Fig. 2(e)) of this semi-infinite configuration with 5 complete unit cells and 2 half cells indicates the suppression of the even modes (blue curves in Fig. 2(b)), leaving only the odd modes. Figure. 2(f) gives the eigen electric field distribution (left panel) and the corresponding electric field phase distribution (right panel) of the bulk mode at K point (indicated by an orange circle). It can be found that such bulk mode features vortex field similar with conventional chiral valley edge modes, i.e., valley-lock. Specially, the electric field of the bulk mode is uniformly distributed throughout the entire structure, which implies the transport channel is the whole volume of the configuration.

Such selectivity of bulk mode can be revealed by the $k·p$ theory. For a semi-infinite ribbon-like configuration, the $k·p$ Hamiltonian near the K valley follows:

$$H(\delta \vec{k}) = v_D(\sigma_x \delta \vec{k}_x + \sigma_y \delta \vec{k}_y) + \Delta m \sigma_z,$$

where $\Delta m$ is the step-mass term characterizing the degree of parity symmetry breaking. $\Delta m$ satisfies both $\Delta m_{up} < 0$ and $\Delta m_{down} > 0$ for the odd modes and

$\Delta m_{up} > 0$ and $\Delta m_{down} < 0$ for the even modes. $\Delta m_{up}$ ($\Delta m_{down}$) is defined as the value of $\Delta m$ in the region upper (lower) the top (bottom) boundary). In our system $\Delta m_{up} = -\Delta m_{down} = -\infty$ is mapped to the PEC boundary case and $\Delta m_{up} = -\Delta m_{down} = +\infty$ corresponds to the PMC case, which implies the parity of the bulk modes is selective by modifying boundary conditions.

## III. RESULTS AND DISCUSSION

To explore the one-way valley-locked modes with large channels, we designed a two-dimensional (2D) PhC as shown in Fig. 3(a). The PhC is composed of 17×8 unit cells with rod radius of r = 2 mm and permittivity $\varepsilon_r$ = 12.4-0.008j. The top and bottom edges adhered with semi-rods are set as PEC boundaries and the other two edges are set as scattering boundaries. The source with left circular polarization (LCP) or right circular polarization (RCP) is accomplished by a four-monopole-antenna array with a phase difference of $\frac{\pi}{2}$ or $-\frac{\pi}{2}$ and settled in the middle of the sample, which is shown in the inset of Fig. 3(f). Figures 3(b) and 3(c) illustrate the electric field distributions at $f$ = 11.78 GHz under the LCP and RCP excitations, respectively. Coinciding with the band structure in Fig. 2(e), the excited EM wave exhibits the one-way property. The propagation direction depends on the source's polarization owing to valley-lock, i.e., LCP excitation along +$x$ axis and RCP along -$x$ axis. It also can be found that the electric field distributes throughout the whole channel, i.e., the width of the configuration with a value of $w = 17a = 4.4\lambda$. Here, $w$ is identified by the number of primitive cells stacked along the +$y$ axis and $\lambda$ is the wavelength of free space at the central frequency $f_0$ = 11.84 GHz. Taking the LCP excitation as an example, Fig. 3(d) gives the integral of energy fluxes Px along the channel width in the whole frequency range of one-way bulk modes. It is clear the energy fluxes at $x$ = 10$a$ (red triangles corresponding to +$x$ axis direction) are much larger than those at $x$ = -10$a$ (blue triangles standing for -$x$ axis direction), showing good unidirectional property. Figure 3(e) plots the magnitude of electric field |Ez| along $x$ = 10$a$ at 11.78 GHz. The electric field expresses a regular

equal-amplitude oscillation versus positions, shows good uniformity along the channel width direction.

Following the parameters above, we fabricated a PhC sample by commercial ceramic ($Al_2O_3$, $\varepsilon_r$ = 12.4-0.008j) rods, as shown in Fig. 3(f). The dielectric rods were machined into a geometric size with radius of r = 2 mm and height of h = 10 mm, except those at the edges crafted into a semi-cylinder. The sample was sandwiched between two parallel metallic plates to simulate a quasi-2D environment. The top and bottom edges of the sample with semi-rods were cladded by two metallic bars, simulating the PEC boundary, and the other two edges were covered by EM-absorbing materials. The slidable detect probe and the fixed feed antenna array, connected to the network analyzer Agilent E8363A, were placed on the top and the bottom metallic plates, respectively. We employed near-field scanning to conduct measurements of the electric field distributions. Figures 3(g) and 3(h) depict the normalized electric field under the LCP and RCP at 12.6 GHz, respectively. It can be found that the experimental results are in good agreement with the simulated ones, substantiating the existence of one-way large channel light transmission within the all-dielectric PhC structure. The frequency deviation is caused by the air layer between the sample and the upper metallic plate [24]. Figures 3(i) and 3(j) present the deduced wavevector in reciprocal space utilizing the electric field within the white dashed box in Fig. 3(g) and 3(h), respectively. The wavevector locates in the vicinity of the K (K') point under the LCP (RCP), consistent with those (Fig. 3(k) and 3(l)) derived from the simulation.

In view of the challenge of simultaneously optimizing three key performance factors, i.e., channel width, operating frequency range, and the uniformity of the waveguides [25–30], we investigate the effect of the channel width on the operating frequency and EM transmission uniformity in our configuration. We fixed the channel length, the size in x direction, at $18a$ and increase the channel width $w$ from $8a$ to $20a$ every $4a$. Figure 4(a) gives the electric field distributions at $x = 15a$ versus different $w$ under the LCP excitation. For the narrow channel, such as $w = 8a$, the field magnitude has little fluctuation even after introducing the materials' loss ($\varepsilon_r$ = 12.4-

0.008j), showing good uniformity. As the transmission channel widens, the average electric field intensity decreases but still maintains good uniformity. This trend is visually depicted in Figs 4(c) ($w = 12a = 6.7\lambda$)) and 4(d) ($w = 16a = 8.9\ \lambda$), where the field in larger channel is weaker than that in the narrower. The electric field distributes uniformly in the region far away from the source at both channels. Furthermore, we studied the operating frequency range versus the channel width $w$. We increased $w$ from $10a$ to $70a$ every $10a$. The results are shown in Fig. 4(b). As the channel expands, the frequency range of the one-way bulk mode shrinks but does not disappear due to the symmetry breaking. The reduction is more significant at smaller $w$ and becomes gentle when $w$ further increases.

Utilizing such one-way valley-locked property, we designed an optical polarization separator robust against multiple corners and an optical tunneling as depicted in Figs. 5(a) and Figs. 5(b), respectively. A line point (blue star) is placed in the middle of the waveguide. Once the wave is excited, the LCP and RCP components are automatically separated. The EM wave carrying the LCP propagates only towards $+x$ axis (right) direction and that with RCP along $-x$ axis (left), showing a good polarization separation. Under the protection of valley-lock, the EM waves with both polarizations show total transmissions all through the channel without backscattering at the sharp corners under an angle of 60°, which increases transmission efficiency. For optical tunneling, it can be divided into two types according to the width of tunneling channel $d$, i.e., concave ($d < 12a$) and convex ($d > 12a$) channels if we fixed the width of the channels connected to tunneling part as $d = 12a$. As shown in Fig. 5(b), for both two cases of $d = 4a$ and $d = 18a$, EM field distributions are nearly the same before and after the tunneling regions, which implies the EM wave transmission is unaffected by the sudden change of the channel width along the propagation path. The good tunneling can also be verified by the transmission curve versus different $d$ as shown in Fig. 5(c). The EM transmission is close to 1 in whichever widths and whatever types, which provides a promising application in optical device.

## IV. CONCLUSIONS

In this work, we proposed an innovative one-way valley-locked waveguide with large channel based on all-dielectric PhC. Employing the mode symmetry, we selectively eliminate a pair of bulk modes, unveiling one-way bulk modes with valley-locked. The electric field distributes uniformly throughout the whole structure bringing a large one-way transmission channel. Both the simulations and experiments affirm the valley-locked nature of these one-way bulk modes, thereby paving a new avenue for optical devices like optical polarization-separator and light tunneling.

## ACKNOWLEDGMENTS

This work is supported by the National Natural Science Foundation of China (Grant No. 12125504, No. 62171215, and No. 62001212), the Priority Academic Program Development of Jiangsu Higher Education Institutions and Jiangsu Provincial Key Laboratory of Advanced Manipulating Technique of Electromagnetic Wave. Y. P thanks the support of the young scientific and technological talents promotion project of Jiangsu Province and Zhongying Scholarship.

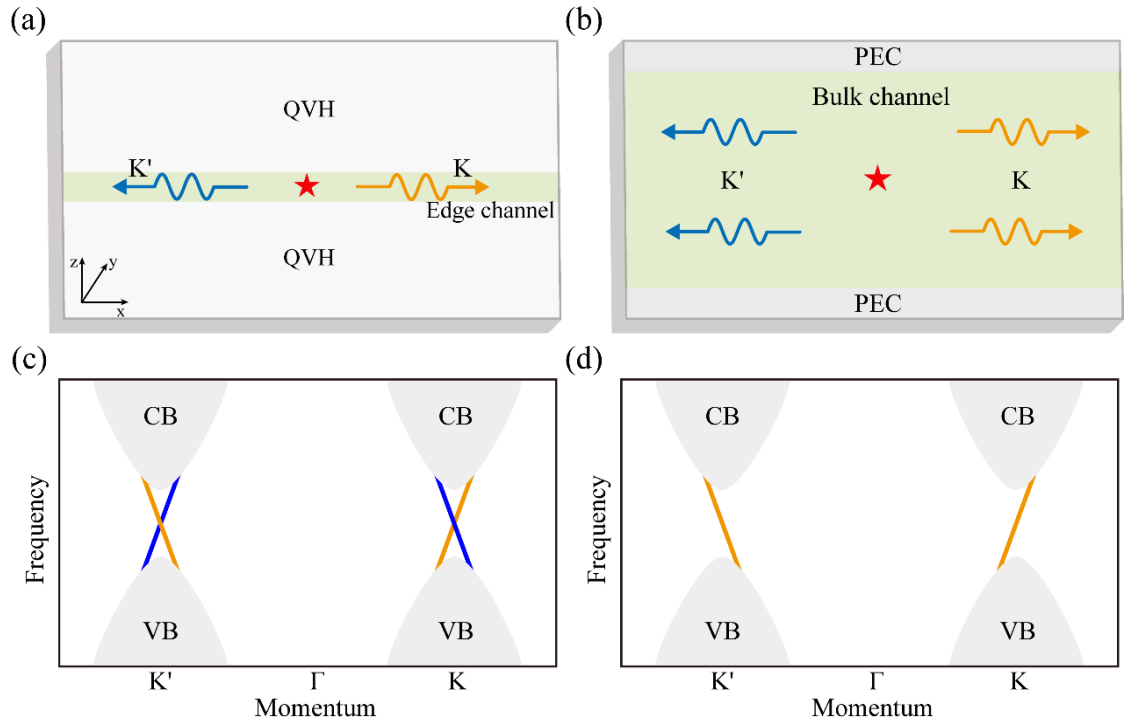

**FIG. 1.** (a) Topological valley-lock edge channel waveguide. The transmission channel is considered as the interface of two topological insulators with quantum valley hall (QVH) phase. (b) One-way valley-lock bulk channel waveguide. The transmission channel completely utilizes the whole structure volume. (c) Schematic of band structure with parity symmetry, where two bands with opposite parity (blue and orange solid lines) intersected at K(K') point. (d) One-way bulk mode with valley-lock.

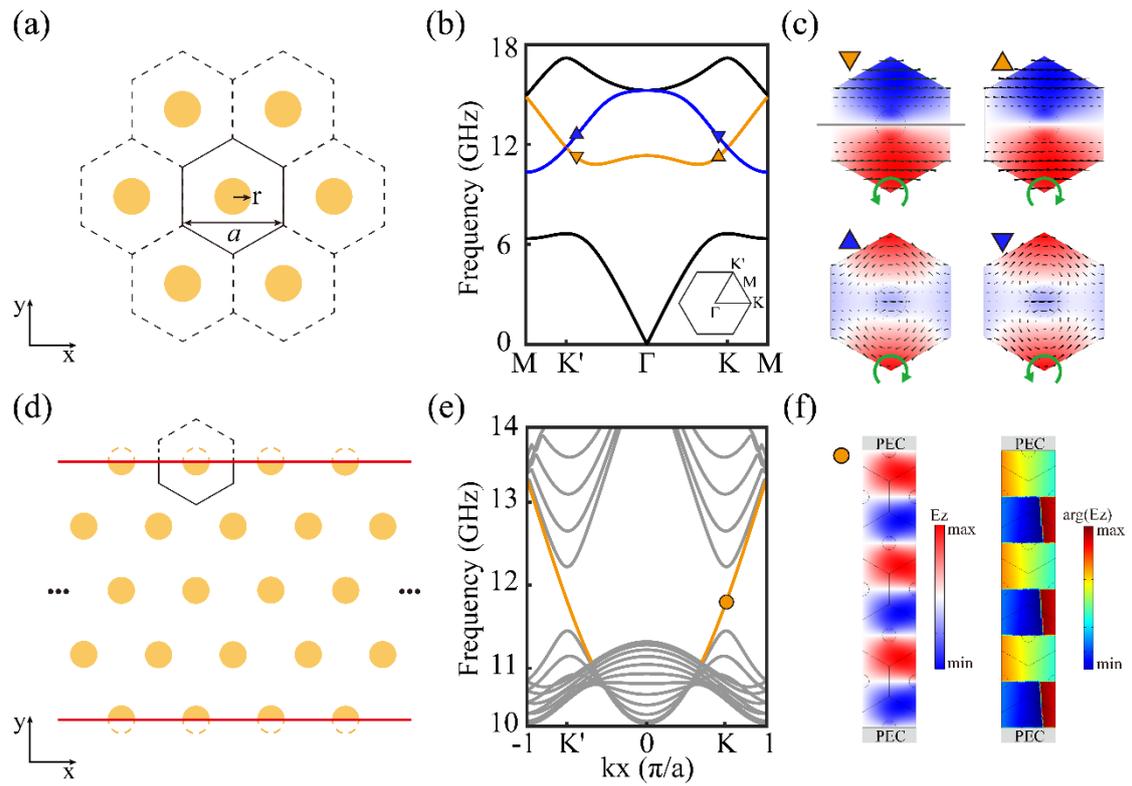

**FIG. 2.** (a) Schematic of hexagonal PhC, where the lattice constant *a* = 16mm and the rod radius r = 2mm. (b) Photonic band structure of hexagonal PhC. (c) Electric field distribution near points K (left panel) and the K' (right panel). The energy flow is indicated by the black arrows. (d) Schematic of semi-infinite ribbon-like supercell. The upper and lower edges along the solid red lines are set as PEC boundary conditions. (e) Band diagram of the semi-infinite ribbon-like supercell. (f) Electric field (left panel) and phase (right panel) distribution at K-point.

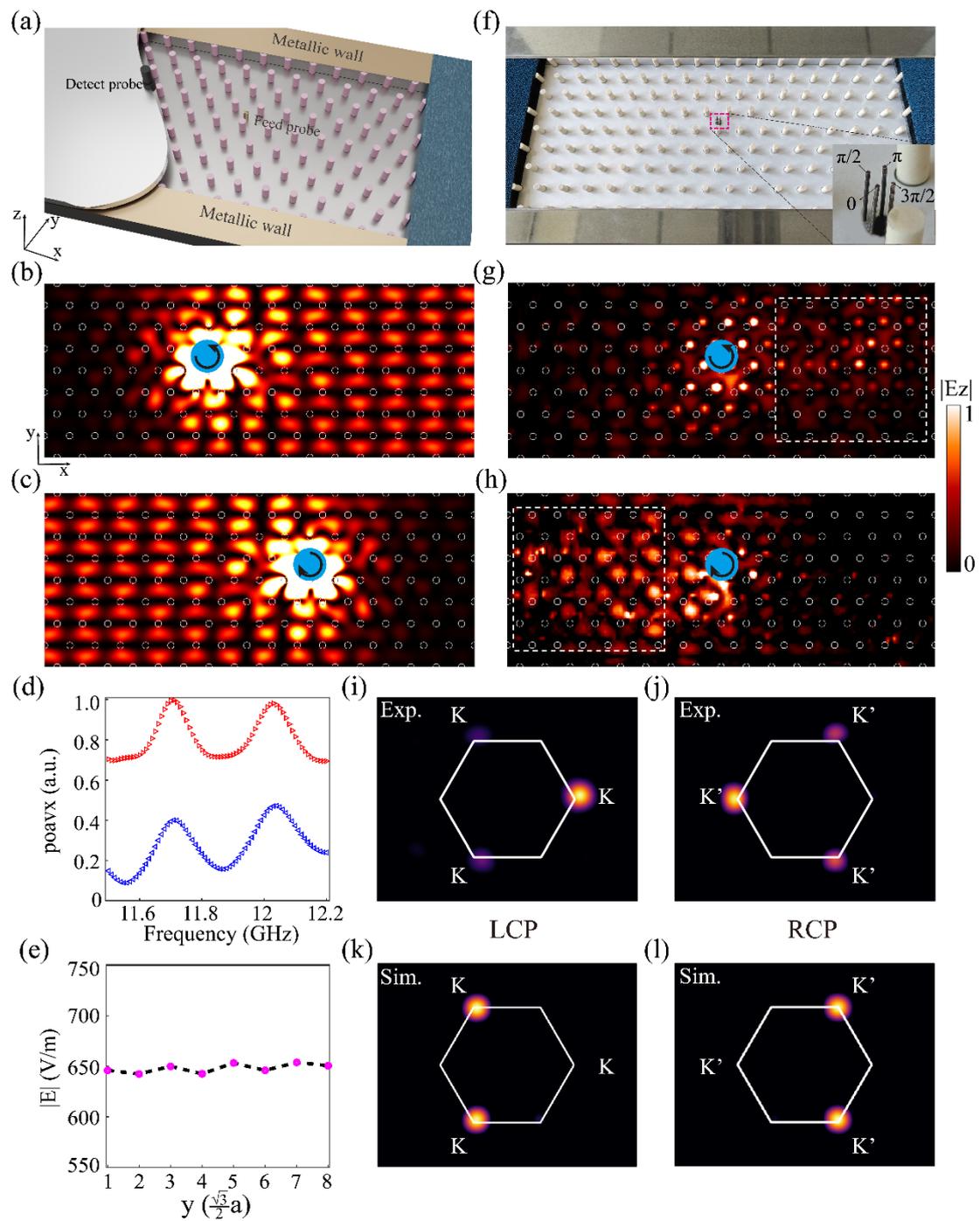

**FIG. 3.** (a) The experimental setups for quasi-2D PhCs. The upper and lower edges adopted metallic cladding and other edges adopted scattering cladding. (b-c) Electric field ($|E_z|$) distributions at 11.78GHz excited by LCP (b) and RCP (c) sources. We set the position of the source to $x = 0$. (d) Energy fluxes along the $+x$ axis (red triangles) and the $-x$ axis (blue triangles) within the operating frequency range. (e) Magnitude of the electric field $|E|$ at 11.78 GHz. (f) Photo of PhC sample. Inset: a four-antenna array

with a phase difference of $\frac{\pi}{2}$ or $-\frac{\pi}{2}$. (i-l) Simulation (i, j) and experimental (k, l) Fourier spectra under LCP and RCP sources obtained from the Fourier transform of the electric field captured within the white dashed box in (g) and (h), respectively.

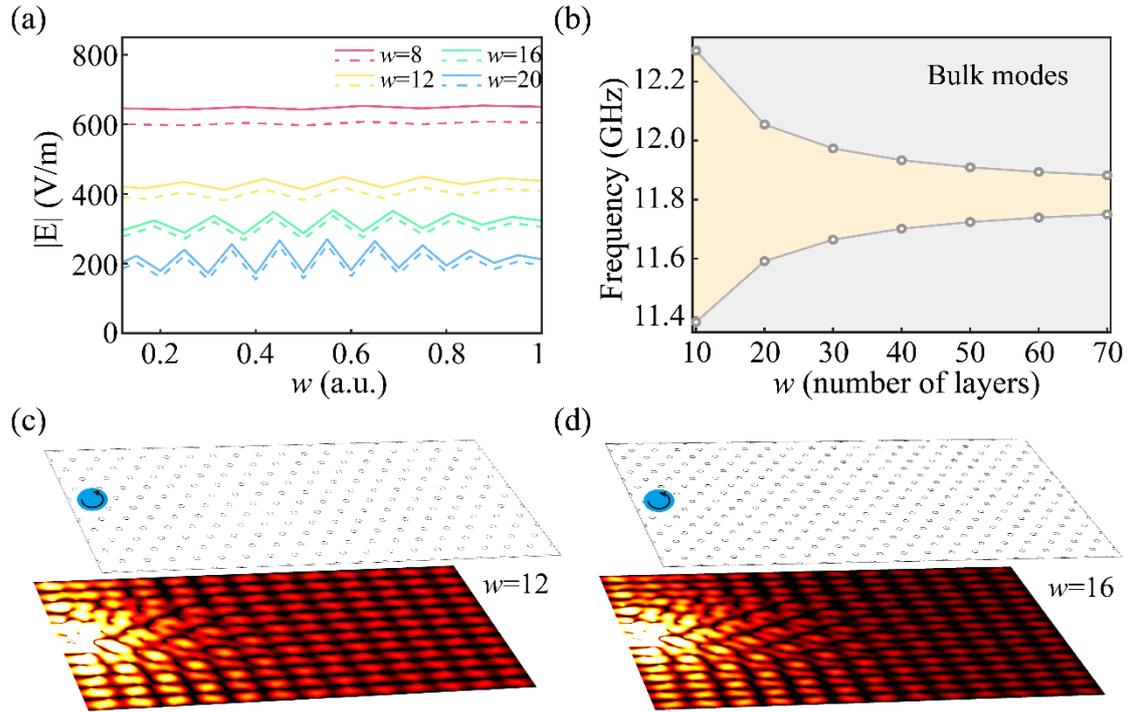

**FIG. 4.** (a) Energy distributions along the y-axis under different $w$. The solid (dotted) lines represented the ideal (lossy) materials. (b) Operating frequency range under different $w$. (c-d) Electric field ($|E_z|$) distribution at $w = 6.7\lambda$ (c) and $w = 8.9\lambda$ (d).

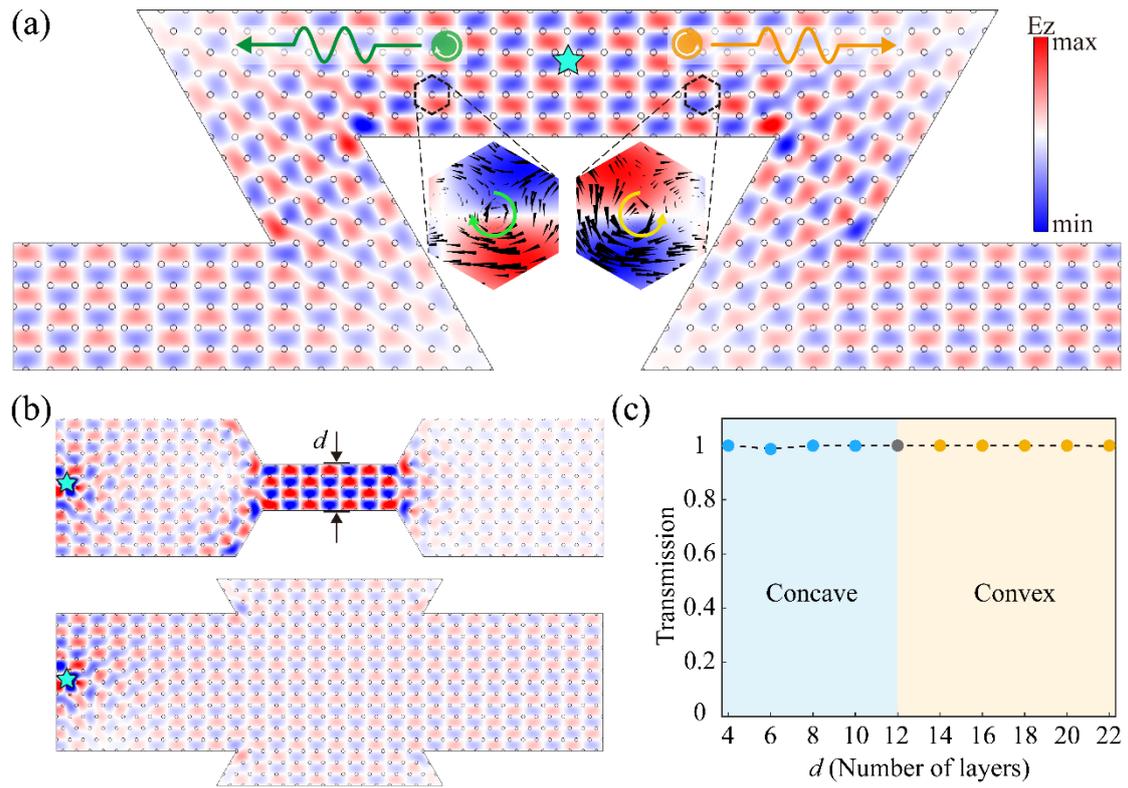

**FIG. 5.** (a) Valley-lock large channel optical polarization separator. The LCP (RCP) component of EM waves propagated along the $x$ (-$x$) direction. (b) Electric field distribution at $d = 4a$ (upper panel) and $d = 18a$ (lower panel). (c) EM transmission with different $d$. The blue and orange regions represented the concave ($d < 12a$) and convex ($d > 12a$) channels, respectively.